\documentclass[12pt,aps,showpacs,preprintnumbers,onecolumn,superscriptaddress,floatfix,nofootinbib]{article}
\usepackage{mathtools}
\usepackage{authblk}
\usepackage{amsmath,amsfonts,amssymb}
\usepackage{graphicx} % Required for inserting images
\usepackage{algorithm}
\usepackage{algorithmicx}
\usepackage{algpseudocode}
\usepackage{xcolor}
\usepackage{comment}
\usepackage{appendix}
\usepackage{hyperref}

\newcommand{\latin}{t}
\newcommand{\greek}{\theta}

\title{Maximum entropy modeling of Optimal Transport: the sub-optimality regime and the transition from dense to sparse networks}

\author[ ]{
Lorenzo Buffa$^{1, 2}$, Dario Mazzilli$^{1,*}$, Riccardo Piombo$^{1}$, Fabio Saracco$^{1}$, Giulio Cimini$^{1,2}$, Aurelio Patelli$^{1}$
}

\affil[1]{Enrico Fermi Research Center, 00184 Rome, Italy}
\affil[2]{Department of Physics and INFN, University of Rome Tor Vergata, 00133 Rome, Italy}
\affil[*]{dario.mazzilli@cref.it}

\begin{document}

\maketitle

\section*{Abstract}
We present a bipartite network model that captures intermediate stages of optimization by blending the Maximum Entropy approach with Optimal Transport. 
In this framework, the network’s constraints define the total mass each node can supply or receive, while an external cost field favors a minimal set of links, driving the system toward a sparse, tree-like structure. 
By tuning the control parameter, one transitions from uniformly distributed weights to an optimal transport regime in which weights condense onto cost-favorable edges.
We quantify this dense-to-sparse transition, showing with numerical analyses that the process does not hinge on specific assumptions about the node-strength or cost distributions.
Finite-size analysis confirms that the results persist in the thermodynamic limit.
Because the model offers explicit control over the degree of sub-optimality, this approach lends to practical applications in link prediction, network reconstruction, and statistical validation, particularly in systems where partial optimization coexists with other noise-like factors.

%%%%%%%%%%%%%% SECTION: INTRODUCTION   %%%%%%%%%%%%%%
\section*{Introduction}
Optimal Transport (OT) theory provides a powerful mathematical framework for transferring one distribution of mass into another at minimal total cost~\cite{OT_general2, Kantorovich2006}. 
Thanks to its elegant formulation and simplicity of arguments, it has found widespread applications in fields such as economics~\cite{OT_economics}, computer science~\cite{OT_image, OT_image2, OT_image3, OT_image4} and biology~\cite{OT_biology, OT_biology2, OT_biology3}. 

\begin{figure*}[htb]
    \centering
    \includegraphics[width=1\textwidth]{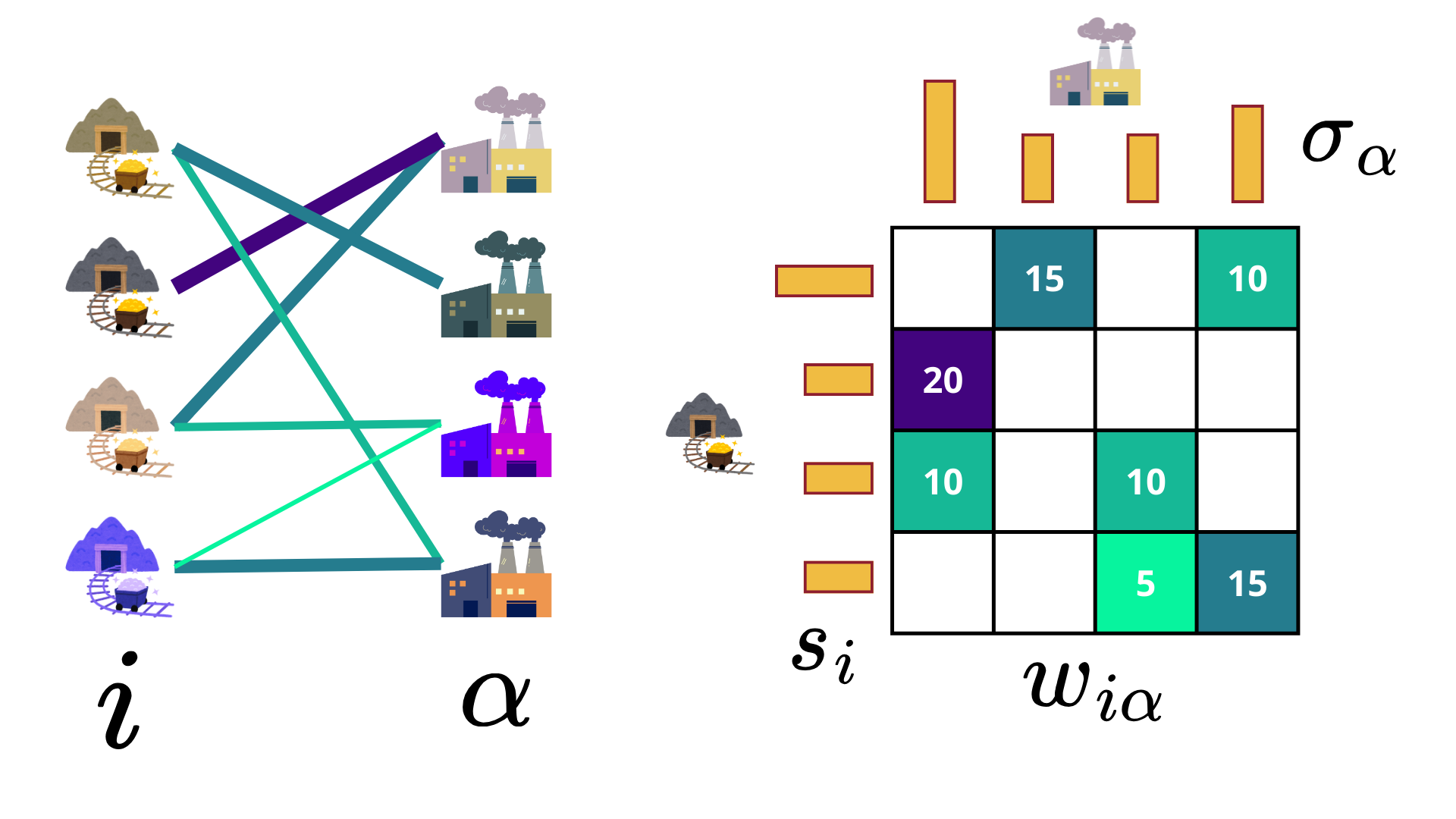}
    \caption{\textbf{Example of Optimal Transport framework.} Optimal Transport solution for a system of $4$ mines that produce coal to be sold to $4$ factories. On the left, the optimal transport plan as a weighted network, where the color and width of each link represent its weight. On the right, the biadjacency matrix $w_{i\alpha}$ of the same bipartite network, where the color of each cell has the same meaning. On the side of the matrix we show the strengths of each node, $s_i$ and $\sigma_\alpha$.}
    \label{fig:OT_cartoon}
\end{figure*}

In the discrete case, OT is well-suited to be described within a bipartite network framework. This setup involves two sets of elements, e.g. $N$ coal mines and $M$ factories, with assigned physical constraints, such as mining capacity and coal necessity respectively, represented by two vectors $\vec{s}$ and $\vec{\sigma}$, as shown in Fig.~\ref{fig:OT_cartoon}.

\noindent
In the discrete case, the transport plan is described by a matrix \(w\). Each element \(w_{i\alpha}\) specifies the amount of mass transported from mine $i$ to factory $\alpha$. To ensure the feasibility of the solution, the plan \(w\) must satisfy the marginal constraints that enforce the preservation of the total mass in each distribution:
\begin{equation}
\label{eq:discrete_constr}
\sum_{\alpha=1}^M w_{i\alpha} = s_i,  \quad \text{and} \quad \sum_{i=1}^N w_{i\alpha} = \sigma_\alpha
\end{equation}
\noindent
These constraints guarantee that the OT solution respects the initial and final distributions: the total amount of mass leaving a source equals its supply, while the total amount received by a destination matches its demand.
The constraints in Eq.(\ref{eq:discrete_constr}) define a polytope $\Gamma$, which represents the set of all matrices $w$ fulfilling the specified conditions.

In this framework, the objective is to minimize the transport cost using a unit cost matrix $C$. Each unit of mass transported from $i$ to $\alpha$ costs $C_{i\alpha}$ and the total cost of a given transportation plan $w$ is simply $ \sum_{i\alpha} w_{i\alpha} C_{i\alpha}$. In the classic OT problem, $s,\sigma,C$ are fixed and given as input to the problem.

The optimization only requires to find the optimal transportation plan $w^*$ such that it minimizes the total cost:

\begin{equation}
\label{eq:discr_OT}
w^* = \underset{w \in \Gamma}{\mathrm{argmin}} \left[ \sum_{i=1}^N \sum_{\alpha=1}^M w_{i\alpha} C_{i\alpha} \right]
\end{equation}

To enhance computational efficiency and stability, Peyré and Cuturi \cite{Cuturi2013, peyre2019computational} introduced an entropic regularization to the OT problem in Eq.(\ref{eq:discr_OT}) that can be solved thanks to the Sinkhorn-Knopp algorithm \cite{Sinkhorn1967}.

Interestingly, the exact solution of Eq.(\ref{eq:discr_OT}) has a peculiar network structure: the optimal transportation/allocation plan is a tree in the bipartite network~\cite{Brualdi2006}. Thus, such optimization would strongly affect the network structure of the system, and many standard problems of Network Theory, such as link prediction and network reconstruction, might have to take into account the possible presence of link preference through OT-like couplings in order to be solved properly. 

Clearly, no real system can achieve perfect optimization and a small degree of sub-optimality is to be expected.
 Fluctuations, noise, competing processes and incomplete information can all cause deviations from the optimal configuration. 

This perspective motivated us to build “noisy OT” models, which aim to bridge the gap between the purely optimal transport plan, characterized by a tree-like structure, and a more diffuse and dense configurations typically observed in empirical data. 

We follow the same approach described in \cite{Koehl19} where the authors built a finite-temperature OT model and characterize the convergence and computational complexity of their solver. We further study the model by defining an order parameter to characterize the transition from dense to sparse states. Moreover, we frame this approach in the field of random network ensembles, where it represents a generalization of pre-existing models, uncovering its potential to be applied as a null model for real data or as tool for network filtering or reconstruction.

We rely on Maximum Entropy (MaxEnt) principle, that provides the most unbiased estimate of a system's microscopic configuration by maximizing Shannon entropy, subject to a set of constraints that determine the values of the observables of interest~\cite{Jaynes1957, Cover2006}. 
In the context of random graphs, the MaxEnt principle prescribes constructing ensembles of graphs that ensure an unbiased representation of all permissible network configurations that remain maximally uninformative beyond the specified constraints~\cite{Cimini2019}.

To formalize this framework, we define \(P(G)\) as the probability of observing a particular network \(G\) in the ensemble \(\mathcal{G}\), and let \(\pi(G)\) be a generic observable whose expected value is constrained to a specified target \(\pi^*\). 
In our setting, each graph \(G\) is a bipartite weighted network uniquely described by its biadjacency matrix \(w(G)\)—an \(N\times M\) matrix whose entries \(w_{i\alpha}(G)\) are positive real numbers that represent the intensity of interactions (weights) between nodes \(i\) and \(\alpha\) in the two distinct layers. 
We also have the constraints on the node strengths \(s_i(G)\) and \(\sigma_\alpha(G)\), which specify how much “mass” or total weight each node can distribute or receive:
\begin{subequations}
    \begin{align}
        & \langle s_i\rangle=\sum_{G\in\mathcal{G}}P(G) s_i(G)=s^*_i, \quad s_i(G)=\sum_\alpha w_{i\alpha}(G),\quad \forall i = 1, \dots, N \\
        & \langle\sigma_\alpha\rangle=\sum_{G\in\mathcal{G}}P(G) \sigma_\alpha(G)=\sigma^*_\alpha, \quad \sigma_\alpha(G)=\sum_i w_{i\alpha}(G),\quad \forall \alpha=1,\dots,M 
    \end{align}
    \label{eq:constraints}
\end{subequations}
In Equations~(\ref{eq:constraints}), the terms \(s^*_i\) and \(\sigma^*_\alpha\) denote fixed values obtained from empirical observations, specifying the expected strengths for each node. 
Incorporating these constraints within the MaxEnt framework ensures that the generated ensemble of graphs aligns with the realistic conditions observed in actual data.

This manuscript is organized as follows: in the Results we show our model of random network and numerical analysis of the transition, in Discussion we comment our findings and potential application and further studies, in Methods we provide the derivation of the model and the details of our numerical algorithm.
%%%%%%%%%%%%%%    SECTION: RESULTS   %%%%%%%%%%%%%%
\section*{Results}
\subsection*{Sub-Optimal Transport Random network model}
A key innovation introduced in this work is the extension of the random graph approach described in the previous section by including of a cost term, which assigns a relative importance to each link through a cost matrix $C_{i\alpha}$. 
Drawing an analogy from physics, this addition is akin to coupling a system to an external field. 
Importantly, the introduction of the cost matrix $C_{i\alpha}$ does not act as a new constraint.
It represents a conceptual shift: instead of maximizing the entropy of the system, we maximize the analogue of a Helmholtz Free Energy $F$, subject to the constraints defined in Equations~\eqref{eq:constraints}. If  $U(G)=\sum_{i,\alpha}w_{i\alpha}(G)\,C_{i\alpha}$ represents the energy of the configuration (the network) $G$, given the disordered external field $C_{i\alpha}$, the $\mathrm{F}$ is defined as
\begin{equation}\label{eq:free_en}
    \mathrm{F}[P] = \mathrm{S}[P] - \beta \mathrm{U}[P],
\end{equation}
where $\mathrm{S}[P] =-\sum_{G\in\mathcal{G}}P(G)\log P(G)$ is the entropy, and $\mathrm{U}[P] = \sum_{G\in\mathcal{G}}P(G)U(G)$.

Solving the optimization problem in Eq.(\ref{eq:free_en}) provides a Boltzmann-like probability distribution:
\begin{equation}
    \label{eq:P_subOT}
    P_{\text{sub\,OT}}(G\vert\beta, \{\latin,\greek\}) = \frac{1}{Z(\beta, \{\latin,\greek\})}e^{-\beta U(G) - \sum_i \latin_i s_i-\sum_\alpha \greek_\alpha \sigma_\alpha}
\end{equation}
Here $\{\latin_i,\greek_\alpha\}$ represent the set of all Lagrange multipliers $\latin_i$ and $\greek_\alpha$ associated with the constraints in Eq.(\ref{eq:constraints}) and $Z$ represents the partition function (i.e. the normalization of the probability distribution).  
It is important to emphasize that the model considers weighted networks where all links are treated as equally significant, as no additional cost is considered. 
In the limit of vanishing $\beta\rightarrow 0$, this model converges to the Bipartite Weighted Configuration Model (BiWCM), discussed in~\cite{Bruno2023}.
When $\beta$ is finite, we label the model \emph{sub OT}. 

The probability distribution in Eq.(\ref{eq:P_subOT}) can be factorized into exponential link probability distributions for each edge $(i,\alpha)$ (see Methods for further details). Thus, it inherently produces a fully connected network, since every pair of nodes \( i ,\alpha \) always has a probability of one of having a link.
Further, the average value of the weight of each link can be computed as
\begin{equation}\label{eq:wia}
    \langle w_{i\alpha}\rangle = \frac{1}{\beta C_{i\alpha} + \latin_i + \greek_\alpha}
\end{equation}
allowing for the construction of an \emph{average} network. 
The values of the Lagrange multipliers of Eq.(\ref{eq:wia}) can be determined by maximizing the associated likelihood function  $\mathcal{L}$ for a fixed value of $\beta$~\cite{Garlaschelli2008}. 

To solve the sub-optimal model numerically, we follow two primary approaches. 
The first is to directly solve the maximum log-likelihood problem through optimization techniques, such as stochastic gradient descent or Adam, available in many libraries for machine learning and neural network models~\cite{NEURIPS2019_bdbca288}.
The second approach involves using the analytical gradient of the log-likelihood to derive a system of nonlinear equations that explicitly include the Lagrange multipliers and the node-strength constraints \(s_i^*\) and \(\sigma_\alpha^*\). 
These equations can then be solved by methods like iterative maps~\cite{vallarano21} or conjugate gradient based algorithms~\cite{saad2003iterative}. 
Both pathways ultimately converge to the same numerical solution for the multipliers since the solution of the problem is unique.
Additional details on the analytical formulation and the numerical procedures for each approach are provided in the Methods.

\begin{figure*}[t!]
    \centering
    \includegraphics[width=1\textwidth]{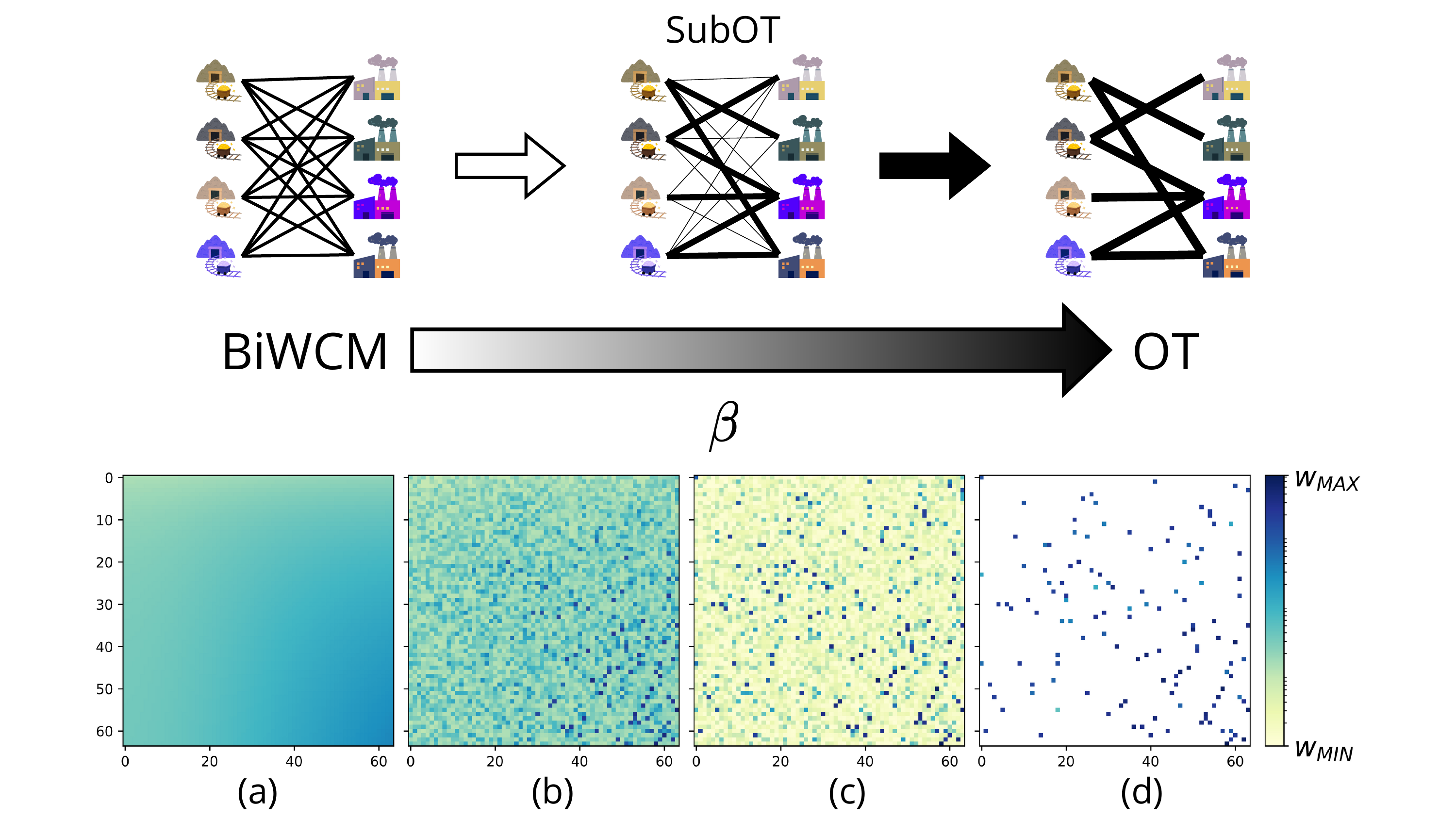}
    \caption{\textbf{Transition between entropy-driven and cost-driven transport regimes.} The upper part of the image is a conceptual illustration of how the transition between different transport plans should manifest as \( \beta \) varies. The OT limit (right) corresponds to the optimal sparse transport solution for moving coal from mines to storage facilities. The center and left illustrations show how introducing an entropic term modifies this transport plan. Initially, it relaxes the strict cost-driven structure, leading to an intermediate sub-optimal regime, where entropy and cost compete in shaping the transport network. Ultimately, it reaches the BiWCM regime, where transport is fully entropy-driven.  
    The lower part of the image provides the quantitative counterpart of the above illustration: it depicts the average transport plan \( \langle w_{i\alpha} \rangle \) in a square network of size \( L=64 \), with uniformly distributed costs $C_{i}\sim U(0,1)$, for different values of the parameter \( \beta \).  
    \textbf{(a)} For low \( \beta \), the weights \( \langle w_{i\alpha} \rangle \) are homogeneously distributed, as mass spreads across all links while satisfying the imposed constraints.  
    \textbf{(b)} As \( \beta \) increases, an underlying structure emerges: some links begin to accumulate larger weights \( \langle w_{i\alpha} \rangle \), causing the network to resemble the tree-like configuration typical of OT.  
    \textbf{(c)} For very large \( \beta \), most of the weight condenses onto the OT tree, shown in panel \textbf{(d)} the exact OT solution.}
    \label{fig:PlansComparison}
\end{figure*}

The parameter $\beta$ tunes the importance of the energy term on the network probability distribution in Eq.(\ref{eq:P_subOT}).
Its impact on the resulting ensemble is illustrated in Fig.~\ref{fig:PlansComparison}, which shows how the expected weights $\{\langle w_{i\alpha} \rangle\}$ evolve as $\beta$ varies. 
What we described in Fig.~1 corresponds to the rightmost illustration in the upper part of Fig.~2, representing the final stage of the OT framework. As shown, the system evolves from the BiWCM, through an intermediate structured state (SubOT), and finally reaches the exact OT plan. 

To be more quantitative, we show in Fig.s~\ref{fig:PlansComparison}a, b, c and d a typical behavior of the average weight matrix during the transition. 

For low $\beta$, our model converges to the distribution of the corresponding BiWCM, where only the entropic term dominates and the external energy vanishes.
In this regime, the weights are distributed homogeneously, according to the constraints, across all links (see Fig.~\ref{fig:PlansComparison}a). 
As expected, their values depend entirely on the prescribed strengths $s^*_i$ and $\sigma^*_\alpha$, resulting in a dense network since no topological constraint is considered. 
The heatmap values are shown on a logarithmic scale, highlighting that in this scenario, the weights cluster tightly around their average values.

As \( \beta \) increases (Fig.s~\ref{fig:PlansComparison}b and c), the weight progressively concentrates on the lowest-cost links, while others become increasingly suppressed. This behavior emerges from our ensemble construction, which maintains fully connected networks with a fixed total weight. In this regime, the system occupies an intermediate state of sub-optimality, where cost minimization and structural constraints compete. 
Consequently, a network corresponding to an intermediate $\beta$ value, between the OT and BiWCM solutions, can be identified as a sub-optimal network relative to the OT solution.

For very large $\beta$, the weights condense in a few links with the lowest value of the cost, and when $\beta \to \infty$, the OT transportation plan is fully recovered (Fig.~\ref{fig:PlansComparison}d) as already discussed in~\cite{Koehl19} (see also Methods in ~\cite{flamary2021pot}).

In conclusion, at varying \( \beta \), we can generate networks that are closer to or further from the OT solution. Therefore our model offers explicit control over the degree of sub-optimality of a network. Since real-world systems are unlikely to perfectly conform to the OT solution, with our method we can naturally explore this intermediate regime of sub-optimality.

In the following sections, we test alternative cost matrices and different strength distributions, providing a broader understanding of the interplay between cost and constraints in shaping the ensemble.

\subsection*{Maximum Spanning Tree and dense-to-sparse transition}
\subsubsection*{Uniform cost and gaussian strength distribution}

\begin{figure}[t!]
    \centering
    \includegraphics[width=\textwidth]{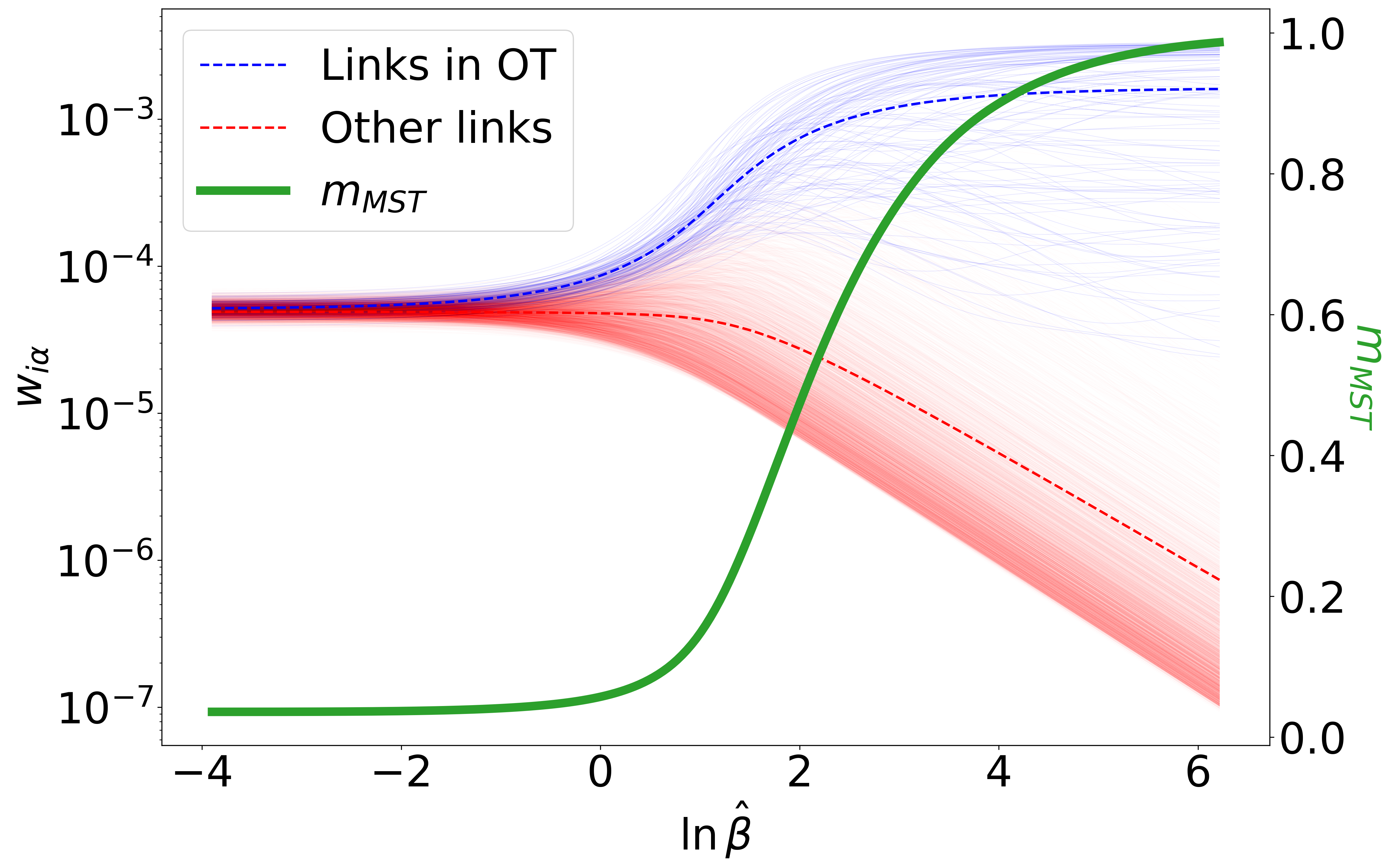}
    \caption{\textbf{The weights $w_{i\alpha}$ as a function of $\ln(\beta)$ illustrate the separation between OT links and the remaining ones.} The figure displays the weights of all links in a $64\times 64$ network with uniformly distributed random costs and Gaussian-distributed strengths. Blue markers represent the weights of links that are part of the OT solution in the $\beta\to\infty$ limit, while red markers correspond to all other links (dashed lines indicate average weights). The green curve shows the corresponding mass of the maximum spanning tree of the network, $m_{MST}$.}
    \label{fig:WeightsChange}
\end{figure}

\noindent
Tracking the evolution of the average weights $\langle w_{i\alpha}\rangle$ with $\beta$, we observe the link separate into two sets between those that form the OT solution and the rest, as shown in Fig.~\ref{fig:WeightsChange}.
At low $\beta$, the two sets are indistinguishable, but get perfectly apart at high $\beta$ values.
Using the terminology of thermodynamics, we refer to the structure at low $\beta$ as the dense phase, since the network is fully connected and any heterogeneity of the weight is driven mainly by the constraints.
Instead, the phase at large $\beta$ corresponds to the sparse phase, where a small number of links (proportional to the number of nodes) get most of the weights.
Since the structure of the OT solution is known to be a spanning tree, we can characterize the degree of sub-optimality at any $\beta$ by looking at the weight associated with the Maximum Spanning Tree (MST) of the average network, defined as the MST mass share:
\begin{equation}
m_{MST}(\beta) = \frac{\underset{(i,\alpha)\in MST}{\sum}\!\!\langle w_{i\alpha}\rangle}{\underset{i,\alpha}{\sum}\,\,\langle w_{i\alpha}\rangle}
\end{equation}

\begin{figure}[t!]
    \centering
    \makebox[\textwidth][c]{\includegraphics[width=1.1\textwidth]{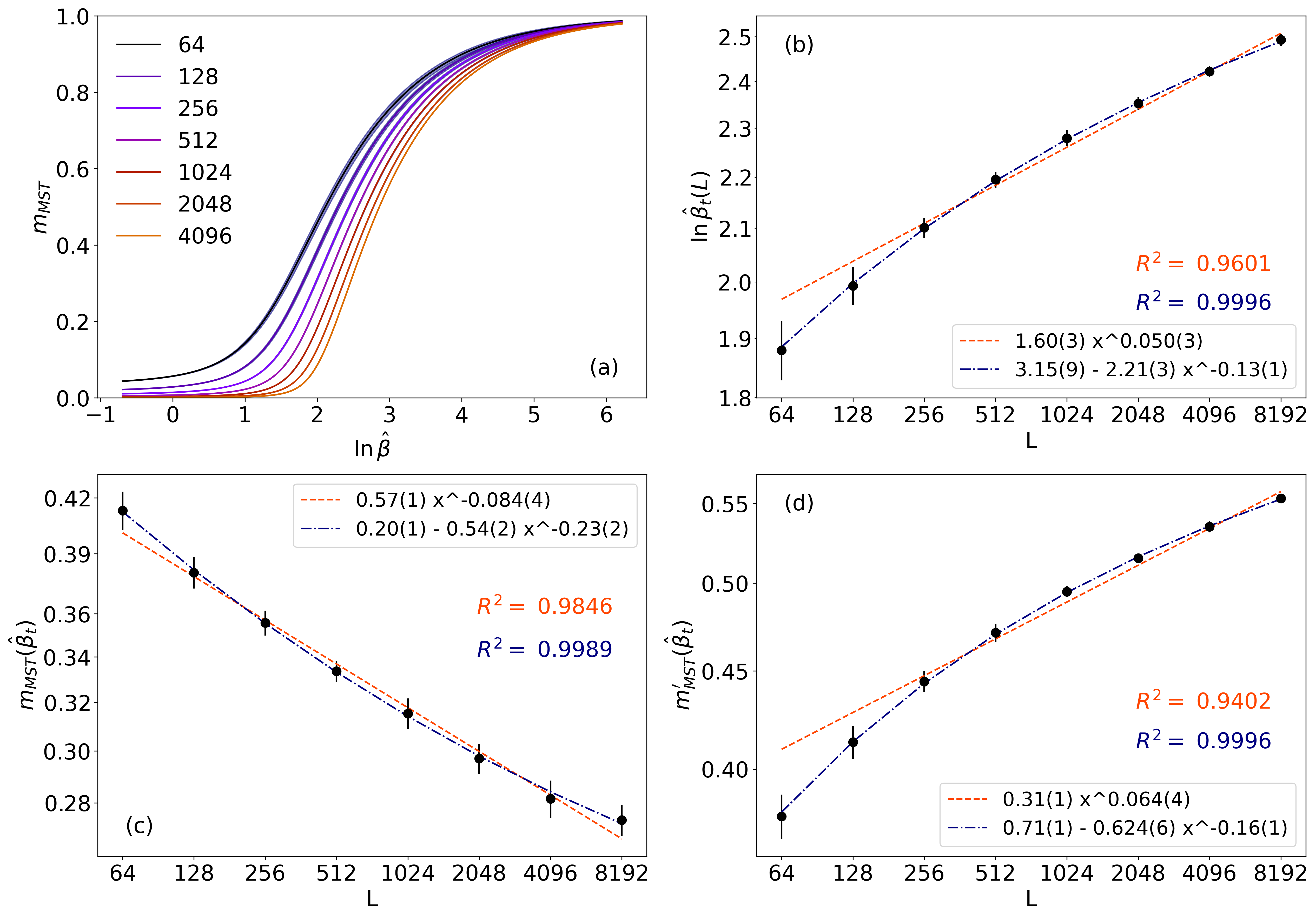}}
    \caption{\textbf{Sub-optimality transition of the model with uniformly distributed costs and Gaussian-distributed strengths.} \textbf{(a)} Behavior of $m_{MST}$ as a function of $\ln\hat\beta$ for different network sizes $L$, where $L$ is the number of nodes in each layer of the bipartite network. Each curve represents the average over at least 100 realizations. \textbf{(b)} Transition values $\ln\hat\beta_\mathrm{t}$ as a function of $L$. \textbf{(c)} Values of the order parameter $m_{MST}$ at the transition point as a function of $L$. \textbf{(d)} Maximum derivative of $m_{MST}$, evaluated at $\hat\beta_\mathrm{t}$. In panels \textbf{(b)}, \textbf{(c)}, \textbf{(d)} orange and blue lines correspond to power-law and bounded power-law fits, respectively. Legends include the best-fit parameters along with the corresponding adjusted $R$-squared values.}
    \label{fig:MSTmass_share}
\end{figure}

The dependence of $m_{MST}$ from $\beta$ is shown in Fig.~\ref{fig:MSTmass_share}a and onward, where we plot it as a function $\ln \beta$. 
This choice is consistent with the mapping proposed in \cite{GabrielliPRE}.
Moreover, because the effective range of the control parameter depends on the total weight of the network, we introduce a rescaled parameter ${\hat\beta} = K \beta$ where \(K=\sum_is_i/L^2\) sets the overall scale of the weights and represents their average value. Thus, when $\ln{\hat\beta}$ appears in any plot, we are referring to this rescaled value.

In our numerical experiments, we consider square (bipartite) matrices of varying linear dimension \(L\) (from \(64\) up to \(4096\)), allowing us to investigate how system size influences the phases of the model.
In Fig.~\ref{fig:MSTmass_share}a we observe that the order parameter \(m_{MST}\) transitions from near-zero values to \(m_{MST}=1\) at increasingly larger \(\hat \beta\) for bigger system sizes, signaling a clear phase transition. Here, the cost matrix $C_{i\alpha}$ has elements drawn from a uniform probability distribution in the range $[0,1]$ and the strengths are Gaussian with mean $5\cdot 10^{-5}L$ and standard deviation $10^{-4}\sqrt{12L}$.
As the system size \(L\) increases, the order parameter $m_{MST}$ grows noticeably steeper, hinting at a saturation in the thermodynamic limit. 
To quantify this behavior, we define the transition value $\hat \beta_{\mathrm{t}}(L) := \arg\max_{\hat \beta}\left[\frac{\partial\,\, m_{MST}(\hat \beta, L)}{\partial\ln\hat \beta}\right]$, following~\cite{LandauBinder2014}, and plot it in Fig.~\ref{fig:MSTmass_share}b.
We then introduce the asymptotic transition value $\beta^*_{\mathrm{t}} \equiv \lim_{L\to\infty} \hat \beta_{\mathrm{t}}(L)$ and examine two possible scenarios: one in which $\beta^*_{\mathrm{t}}$ diverges with the system size (unbounded power law), and another in which it converges to a finite limit (bounded power law). 
While fitting the diverging case with a power-law growth yields a reasonably strong adjusted $R^2$ of about $0.95$, the saturating fit aligns almost perfectly with the data, reaching an adjusted $R^2$ close to 1. 
Consequently, our results suggest that $\ln \beta^*_{\mathrm{t}}\simeq 3.15(9)$\footnote{We use a shorthand notation to express uncertainties. The digits in parentheses represent the error in the last digits of the value.} is finite. 

Fig.~\ref{fig:MSTmass_share} highlights additional features of the transition in our numerical simulations, suggesting that it is non-critical in the sense that no divergences appear in other observables. 
In panel (c), for instance, the order parameter evaluated at the transition point seems to converge to a finite value.

Even more telling is the fact that the first derivative of the order parameter remains finite—no divergence is observed—implying that in the thermodynamic limit, the slope of the transition curve remains bounded.

Despite this non-critical character, the transition can still be usefully characterized by examining the behavior of $m_{MST}$ on either side of $\hat\beta_{\mathrm{t}}$. To that aim we call $\hat \beta_<$ ($\hat \beta_>$) the values of $\hat \beta$ before (after) $\hat\beta_{\mathrm{t}}$.
For $\hat\beta_{<} \ll \hat\beta_{\mathrm{t}}$, the expected weights $\langle w_{i\alpha}\rangle$ are nearly uniform, consequently, $m_{MST}$ decreases as a power law, acting roughly as $m_{MST}(\ln\hat\beta_{<}) \simeq 2/L$. 
In contrast, for $\hat\beta_{>} \gg \hat\beta_{\mathrm{t}}$, $m_{MST}$ approaches a size-independent value that depends on $\hat\beta$, ultimately tending toward 1 as $\hat\beta \to \infty$. 
Thus, in the thermodynamic limit, the system remains “decoupled” from the cost function for $\hat\beta < \hat\beta_{\mathrm{t}}$, only to move into a regime where it systematically converges to an OT–like solution for $\hat\beta > \hat\beta_{\mathrm{t}}$.

\begin{figure}[t!]
    \centering  
    \includegraphics[width=1\textwidth]{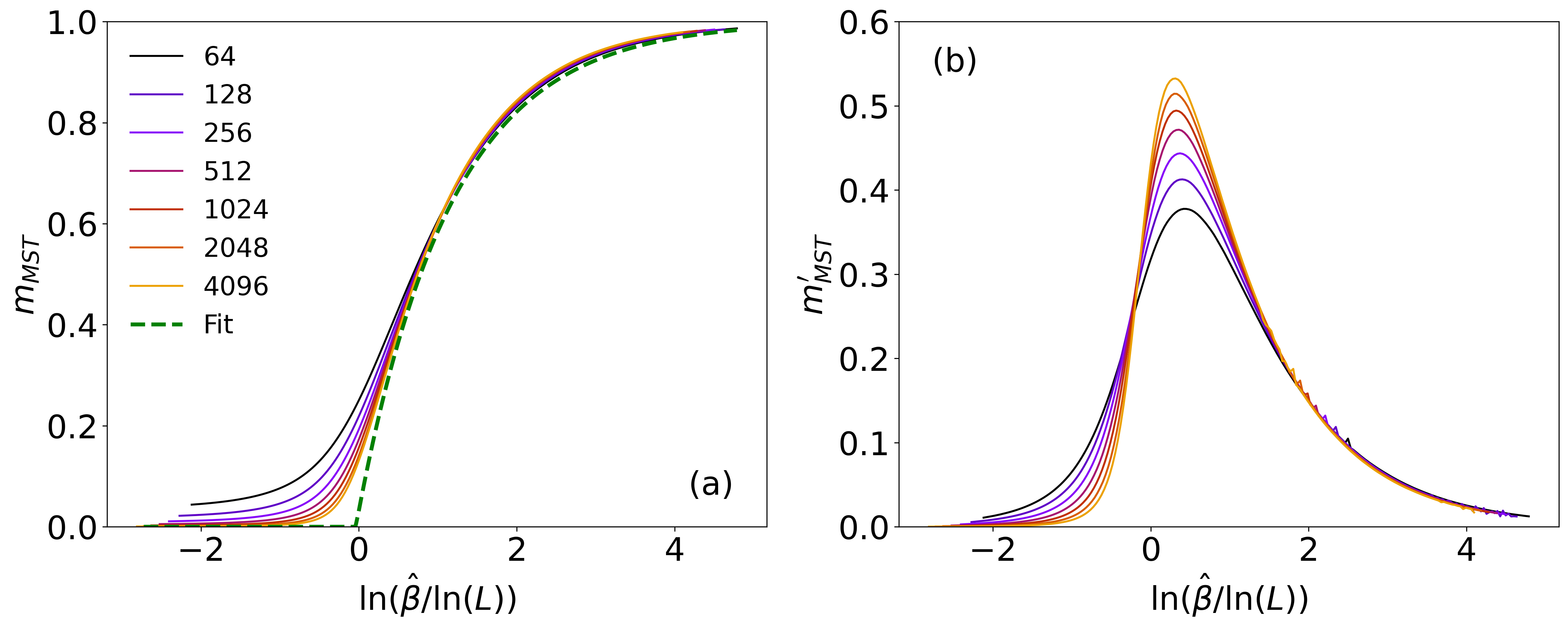}
    \caption{\textbf{The transition properties of the model with a rescaling of the $\beta$ parameter.} \textbf{(a)} The $m_{MST}$ using the rescaling of the control parameter from $\ln\hat\beta$ to $B=\ln(\hat\beta/\ln(L))$. The curve at different sizes overlap in a region where the value of the order parameter indicates the system is already transitioned to the dense phase. The limit curve for the order parameter (dashed in green) is $1-\exp(-\alpha(B-B_C))$ for $B>B_C$ and $0$ for $B<B_C$ (the fitted values are $B_C=-0.039(7)$ and $\alpha=0.846(7)$). The fit has been executed using the $L=4096$ data, which would be our closest approximation of the limit curve. \textbf{(b)} The derivatives of $m_{MST}$ with respect to $B$. The proposed rescaling of the control parameter affects the finite size scaling.}
    \label{fig:BKTmassshare}
\end{figure}
For the sake of completeness, we test our results with scaling of transitions that are not critical.
We propose a new control parameter $B=\ln(\hat\beta/\ln(L))$ that apparently allows for a curve collapse of the $m_{MST}$ and its derivative, as shown in Fig.~\ref{fig:BKTmassshare}. 
The curves collapse is consistent with the observation of a limit curve for the order parameter (dashed in green) that is proportional to $1-\exp(-\alpha(B-B_C))$ for $B>B_C$. The transition does not present a divergence in the first derivative of the order parameter, but shows a slope of $\alpha=0.846(7)$ after the "critical" point $B_C$. Indeed, the exponential fit function is maximum in $B_C$, as expected by empirical behavior.

\subsubsection*{Different cost and strength distributions}
Most of the preceding analyses assumed a specific setup in which the cost functions were uniformly sampled between 0 and 1, and the node strengths were drawn from a Gaussian distribution. 
However, to gain a more comprehensive understanding of the dense-to-sparse transition and to test the robustness of our algorithm, we also examined the model under various cost and strength distributions — showing that our results are not limited to the specific case of uniform costs and Gaussian strengths.
This step is essential for assessing whether the key features of the transition, particularly its nature, are robust to changes in these distributions, and thus whether the model can reliably capture real-world scenarios, where costs and constraints can be highly heterogeneous and only partial information about them may be available.

We first examined the case where the cost remains uniformly distributed, but the node strengths follow a power-law distribution. 
Specifically, we used a probability density function $p(x) \sim x^{-4}$, although similar numerical experiments we tested with other power-law exponents revealed no notable qualitative differences. 
This setup does demand more computational resources because obtaining solutions across different $\hat{\beta}$ values can sometimes require fine-tuning of the simulation parameters.
We also examined the case of uniform distributed costs, but with node strengths following a truncated log-normal distribution, with parameters $\mu=1$ and $\sigma=0.5$ and with a lower bound of $1$ in the support. This has allowed for a similar effect on the strength distribution as the power-law distribution, but with a minor computational effort.

\begin{figure}[t!]
    \centering  
    \makebox[\textwidth][c]{    \includegraphics[width=1.2\textwidth]{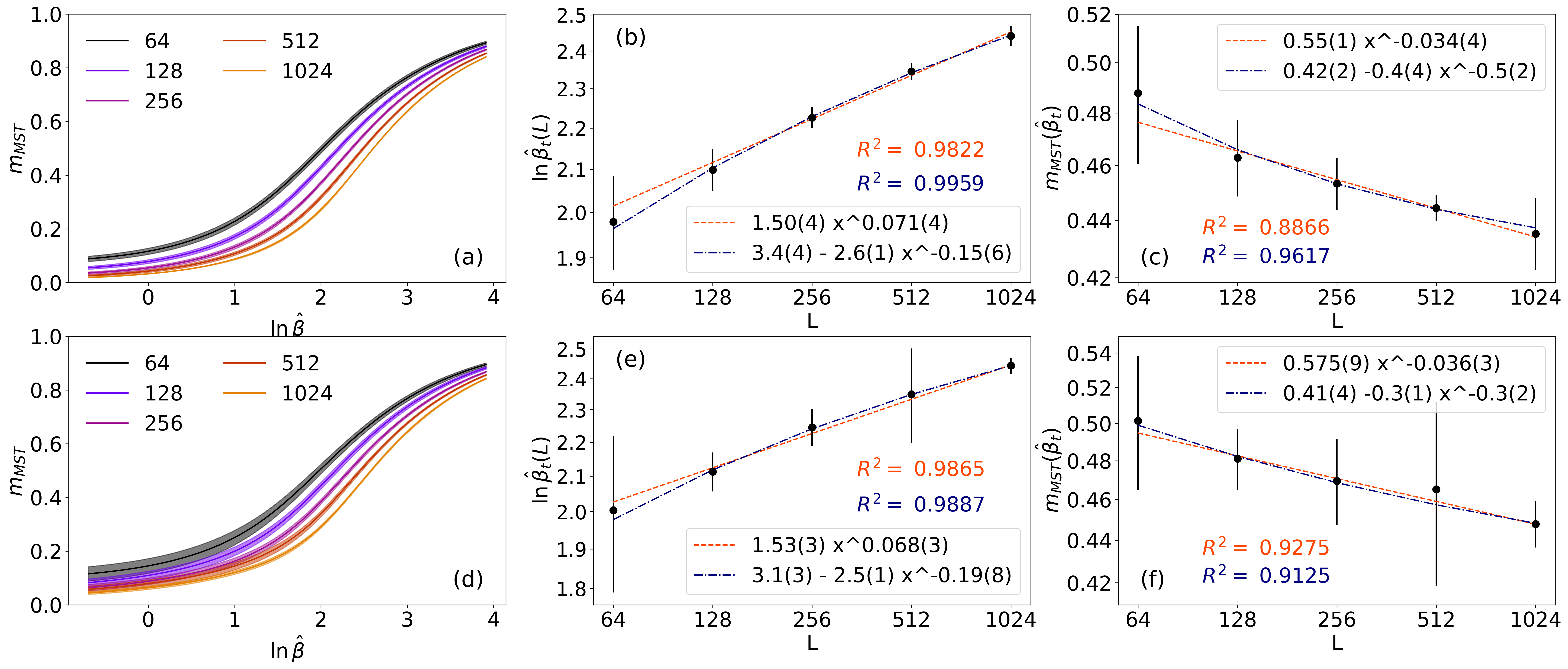}    }
    \caption{\textbf{The transition's properties of the model in the case of Log-Normal-distributed and Power-Law-distributed strengths.} The top and bottom rows refer to the Log-Normal and Power Law distributions, respectively. \textbf{(a)} and \textbf{(d)} $m_{MST}$ as a function of $\ln\hat\beta$, for different sizes $L$, where $L$ is the number of nodes in each layer of the bipartite network. Each line is the average of at least 100 realizations. \textbf{(b)} and \textbf{(e)} The transition values $\ln\hat\beta_\mathrm{t}$ (dots) with respect to $L$. \textbf{(c)} and \textbf{(f)}, maximum derivative of $m_{MST}$, evaluated at $\hat\beta_\mathrm{t}$. The orange and blue lines in panels \textbf{(b)}, \textbf{(c)}, \textbf{(e)}, \textbf{(f)} are a power law and a bounded power law fits. Legends include the best-fit parameters along with the corresponding adjusted $R$-squared values.}
    \label{fig:powerlawstrength}
\end{figure}
The results, illustrated in Fig.~\ref{fig:powerlawstrength}, suggest that the transition is qualitatively similar to the case where the strengths are Gaussian-distributed. 
Moreover, estimating the transition point in the thermodynamic limit yields \(\ln \beta_\mathrm{t}^* \sim 3.1(3)\) for power-law distributed strengths and \(\ln \beta_\mathrm{t}^*\sim 3.4(4)\) for log-normal strengths. Both estimates are consistent with the gaussian-distributed strengths case (see Fig.~\ref{fig:MSTmass_share}b), where we found \(\ln \beta_\mathrm{t}^* = 3.15(9)\).
Although the behavior of the order parameter and its derivative at the transition point is not sufficiently sensitive to decisively distinguish between divergent or saturating fits, our findings are broadly in line with those from the previous setup.

In the second scenario, we vary the cost distribution while retaining Gaussian-distributed node strengths. 
Because costs must be bounded from below, we cannot simply use standard distributions like the unbounded Gaussian often employed in random matrix theory. 
Moreover, as the system size \( L \) increases, uniform distributions tend to produce many cost values that are extremely close — with average spacing scaling as \( L^{-2} \) — which can cause numerical issues, especially in the lower end of the spectrum.
In particular, for large $L$, small cost gaps may create shallow gradients in the Free Energy landscape for configurations that are not true minima; this complication impedes a straightforward transition to the sparse phase. 
Simply rescaling the uniform distribution does not resolve this problem, since it amounts to a trivial shift of the control parameter.
\begin{figure}[t!]
    \centering  
    \includegraphics[width=1\textwidth]{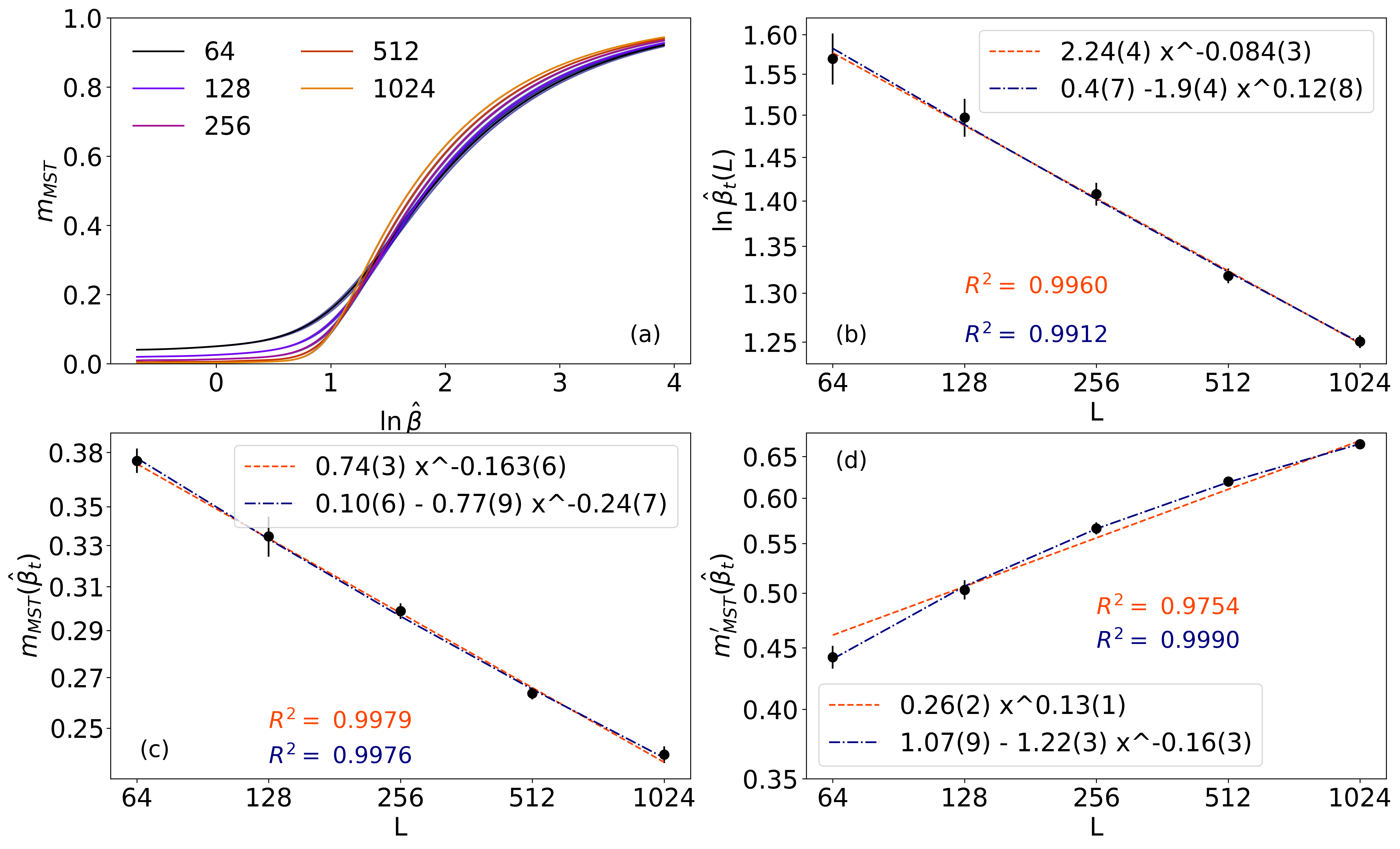}
    \caption{\textbf{The transition's properties of the model in the case of Beta-distributed cost matrices.} \textbf{(a)} $m_{MST}$ as a function of $\ln\hat\beta$, for different sizes $L$, where $L$ is the number of nodes in each layer of the bipartite network. Each line is the average of at least 100 realizations. \textbf{(b)} The transition values $\ln\hat\beta_\mathrm{t}$ (dots) with respect to $L$. \textbf{(c)} The value of the order parameter at the transition point with respect to the layer size $L$. \textbf{(d)} Maximum value of $m_{MST}^\prime$, evaluated at $\hat\beta_\mathrm{t}$. The blue line and the red line are a power law and bounded power law fits respectively. Legends include the best-fit parameters along with the corresponding adjusted $R$-squared values.}
    \label{fig:rootcost}
\end{figure}

Instead, we transform the uniform distribution via a power-law function that depends on $L$:
\begin{equation}
    C_{i\alpha} \;=\; x^{\tfrac{1}{\log_2(L)}}, \quad x \in [0,1],
\end{equation}
so that each new cost variable follows a Beta distribution whose parameters are $\log_2(L)$ and $1$. 
This transformation mainly affects the lower end of the spectrum distribution, preserving a finite gap between minimal cost values as $L$ grows.
As expected, the resulting behavior differs from the simple uniform-cost case, owing to the explicit dependence of the cost distribution on $L$. Although the infinite-size limit of the $m_{MST}$ curves still vanishes below $\ln\hat\beta_{\mathrm{t}} \approx 0.4(7)$ and then rises to $1$ above it, the finite-size curves exhibit intersection points that shift the transition to smaller values of $\hat{\beta}$. The critical point $\hat{\beta_t}$ goes to zero as the system's size increases (panel b), as well as the value of the order parameter $m_{MST}(\hat{\beta_t})$ (panel c), although the two fit, orange and blue lines, have very similar adjusted $R^2$. Even in this setting, the asymptotic value of $m'_{MST}(\hat{\beta_t})$ shows a finite value (panel d).

%%%%%%%%%%%%%% SECTION: DISCUSSION   %%%%%%%%%%%%%%
\section*{Discussion} 
The model presented in this study is grounded in an information-theoretic framework similar to maximally entropic null models, but with an added cost term that acts like an external field, hence modifying the standard Entropy into a Free Energy. 
Our model offers a systematic way to analyze intermediate stages of optimization in bipartite networks, capturing both the uniform \emph{dense} phase, where interaction weights are distributed broadly, and the \emph{sparse} phase, where weights condense onto a minimal set of links, converging toward the spanning tree typical of OT solutions. 
This framework is particularly relevant for domains in which mutualistic networks exhibit strong link preferences, concentrating most of the total interaction mass on a select few edges. 
Within such systems, one can leverage the MST mass share, $m_{MST}$, as an order parameter that tracks the system’s transition between these two extremes.

By tuning the control parameter $\beta$, the model smoothly moves from a maximum-entropy description of bipartite weighted configuration models—where costs play no role—to a near-OT state in which low-cost edges dominate. 
Notably, the transition between dense and sparse regimes does not appear to be \emph{critical} in the statistical physics sense, as we observe no divergences in the conventional thermodynamic indicators. 
Instead, the transition is signaled by the rise of $m_{MST}$, which vanishes below a threshold $\beta_{\mathrm{t}}$ and grows steadily above it, eventually saturating as $\beta$ becomes large. Finite-size analysis confirms the results are robust in the thermodynamic limit of large system size.
Numerical studies consistently confirm this picture, demonstrating that it persists across various strength distributions (e.g., Gaussian and power-law) and cost matrices, including those whose values are adjusted to address finite-size effects. 
These results underscore the robustness of the transition, suggesting that the underlying mechanism of weight condensation onto cost-efficient links holds broadly rather than relying on a specific distributional choice.

Interestingly, although the MST structure itself is central to defining $m_{MST}$, the observed transition is not primarily driven by a topological rearrangement of links. 
Instead, it reflects a smooth redistribution of weights that eventually “locks in” a small number of cost-favorable edges. 
As $\beta$ decreases, these edges lose their advantage, causing weights to diffuse but not necessarily maintaining the OT spanning tree topology. 
In many practical systems, such sub-optimal configurations emerge from a combination of constrained optimization and other “noise-like” processes that deviate from a complete OT solution. 
Hence, our model underscores how a partial optimization mechanism, operating under known or estimated cost functions, can naturally coexist with the myriad factors that hold real networks short of the perfectly optimized state.

From an applied perspective, this framework can be used as a null model for tasks such as link prediction, network reconstruction, and statistical validation. 
If the cost matrix is known or can be reliably estimated, one can deduce the degree of optimization by measuring $\beta$. 
Conversely, if $\beta$ can be inferred—e.g., via external data or by calibrating the MST mass share—then the cost structure can be approximated or constrained accordingly. 
Such capabilities are particularly valuable in fields like ecology (e.g., plant–pollinator interactions) and economics (e.g., trade networks), where sub-optimal strategies may be adaptive responses to uncertain environments or system-level trade-offs. 
By providing a lens on how \emph{sub-OT} mechanisms play out in realistic settings, this model offers a unifying perspective that connects maximal-entropy bipartite null models with the classical OT framework—ultimately helping us quantify, interpret, and predict the patterns found in complex bipartite networks.

%%%%%%%%%%%%%% METHODS %%%%%%%%%%%%%%
\section*{Methods}

\subsection*{Derivation of the Maximum Free Energy Ensemble}
We analyze a canonical ensemble $\mathcal{G}$ of $N \times M$ undirected, weighted bipartite graphs $G$. Each one is represented by a biadjacency matrix whose elements are continuous real numbers representing edge weights $w_{i\alpha}(G) \in (0,\infty)$. The ensemble is equipped with a probability measure $P(G)$, which will depend on a set of parameters, the Lagrange multipliers $\{t,\theta\}$ and the coupling term $\beta$, characterizing the model. To quantify how well this ensemble represents a given real-world network~\footnote{Networks and graphs are interchangeable terms in this work.}, we define the expected value for any observable $\pi(G)$ as:

\begin{equation}
    \langle \pi \rangle = \sum_{G \in \mathcal G} \pi(G)\,P(G)
\end{equation}

\noindent
Therefore, the ensemble's expected value of $\pi$ weights each graph’s value $\pi(G)$ by the probability of observing that graph. Our goal is to determine the parameters $\{t^*,\theta^*\}$ ensuring that the ensemble’s expected statistics match empirical values drawn from an observed network $G^*$ at each value of $\beta$.

In this work, we impose local constraints on node strengths on each layer, and we call those fixed empirical values as $s_i^*$ and $\sigma_\alpha^*$. However, we could have also focused on the degree distribution of each node or considered both aspects simultaneously~\cite{vallarano21,Cimini2019}. Our choice leads to the following constraints, which are already presented in the main text in Equations.~\eqref{eq:constraints}:
\begin{eqnarray}
\label{eq:constr1}
&\langle s_i\rangle=\sum_{G\in\mathcal{G}}P(G)\,s_i(G) = s^*_i \quad \,\,\forall i = 1,\dots, N \\
\label{eq:constr2}
&\langle \sigma_\alpha\rangle= \sum_{G\in\mathcal{G}}P(G)\,\sigma_\alpha(G) = \sigma^*_\alpha \quad \forall \alpha = 1,\dots, M
\end{eqnarray}
\noindent
where
\begin{equation}
s_i(G) = \sum_{\alpha} w_{i\alpha}(G) \qquad
\sigma_\alpha(G) = \sum_{i} w_{i\alpha}(G)
\end{equation}
are the strength variables of the two layers. Obviously, we are also implying the $\sum_{G\in\mathcal{G}}P(G)=1$ constraint that ensures the correct probability normalization. 

Due to the presence of a cost matrix, we seek a maximum-free energy probability measure consistent with Equations~(\ref{eq:constr1}) and (\ref{eq:constr2}). Therefore, as anticipated in the main text, we introduce a free energy functional defined as:
\begin{equation}\label{eq:free_en2}
    \mathrm{F}[P] = \mathrm{S}[P] - \beta \mathrm{U}[P],
\end{equation}
where $\mathrm{S}[P] =-\sum_{G\in\mathcal{G}}P(G)\log P(G)$ is the entropy, and $\mathrm{U}[P] = \sum_{G\in\mathcal{G}}P(G)U(G)$.

We obtain the expression of the probability distribution $P(G)$ by maximizing $\mathrm{F}$ subject to the constraints in Equations~(\ref{eq:constr1}) and (\ref{eq:constr2}). In practice we have to perform, and then set to zero, the functional derivative with respect to $P(G)$ of the following expression:
\begin{eqnarray}
\nonumber
&& -\sum_{G\in\mathcal{G}}P(G)\log P(G) + \gamma\left(1-\sum_{G\in\mathcal{G}}P(G)\right)-\beta \sum_{G\in\mathcal{G}}P(G)\,U(G) + \\
\label{eq:to_derive}
&&\!\!\!\!\!\!\!\!+\sum_i \latin_i \left(s_i^*-\sum_{G\in\mathcal{G}}P(G) \,s_i(G)\right)
+\sum_\alpha \greek_\alpha \left(\sigma_\alpha^*-\sum_{G\in\mathcal{G}}P(G) \,\sigma_\alpha(G)\right)
\end{eqnarray}
\noindent
where $\latin_i$ and $\greek_\alpha$ are the Lagrange multipliers associated with each node in both layers and $\gamma$ is the Lagrange multiplier linked to the normalization constraint. The derivative of Eq.(\ref{eq:to_derive}) gives:
\begin{equation}
    -\log P(G)-1-\gamma-\sum_i \latin_i \,s_i(G)-\sum_\alpha \greek_\alpha \,\sigma_\alpha(G)-\beta\, U(G)=0
\end{equation}
\noindent
Therefore, the probability measure that characterize the ensemble is:

\begin{equation}
    P_{\text{sub OT}}(G \!\mid\! \beta, \{t,\theta\}) =e^{-(\gamma+1)}e^{-\mathcal{H}(G \mid \{t,\theta\})-\beta\, U(G)}.
\end{equation}

\noindent
We enclose all the contributions relative to the Lagrange multipliers in the following quantity:
\begin{equation}
\label{eq:dot_prod}
\mathcal{H}(G \!\mid\! \{t,\theta\})\equiv \sum_i \latin_i s_i(G)+\sum_\alpha \greek_\alpha \sigma_\alpha(G)=\sum_{i,\alpha} w_{i\alpha}(G) (\latin_i+\greek_\alpha)
\end{equation}
\noindent
The partition function is:
\begin{equation}
Z\,(\beta, \{t,\theta\}) = e^{1+\gamma}=\sum_{G\in\mathcal{G}}e^{-\mathcal{H}(G \mid \{t,\theta\})-\beta\, U(G)},
\end{equation}

\noindent
We can rewrite the expression in a more elegant form, resulting in a Boltzmann-like probability distribution for graphs:

\begin{eqnarray}
P_{\text{sub OT}}\,(G \! \mid \!\beta, \{t,\theta\}) &&\equiv \frac{e^{\displaystyle-\mathcal{H}(G \!\mid\!\{t,\theta\}) - \beta \,U(G)}}{Z\,(\beta, \{t,\theta\})}\\
\nonumber
&&= \frac{1}{Z\,(\beta, \{t,\theta\})}\,\exp{\left[-\beta \,U(G) - \sum_i \latin_i s_i(G)-\sum_\alpha \greek_\alpha \sigma_\alpha(G) \right]}\\
\nonumber
&&=\frac{1}{Z\,(\beta, \{t,\theta\})}\,\exp{\left[-\sum_{i,\alpha} w_{i\alpha}(G)\, \left(\beta \,C_{i\alpha} + \latin_i + \greek_\alpha\right)\right]}
\end{eqnarray}
\noindent
The explicit expression of the partition function
\begin{equation}
Z\,(\beta, \{t,\theta\}) = \sum_{G \in \mathcal G} \exp{\left[-\sum_{i,\alpha} w_{i\alpha}(G)\, \left(\beta \,C_{i\alpha} + \latin_i + \greek_\alpha\right)\right]}
\end{equation}

\noindent
can be analytically computed by considering the symbolical sum over the ensemble of graphs as an integral over the real values of the weights $w_{i\alpha}$:
\begin{equation}
    \begin{split}
        Z\,(\beta, \{t,\theta\})&=\int_0^{+\infty}\prod_{i,\alpha} dw_{i\alpha}\,e^{- \mathcal{H}(G \mid \{t,\theta\})-\beta\,U(G)}=\\
        &=\int_0^{+\infty}\prod_{i,\alpha} dw_{i\alpha}\,e^{- \left(\sum_i\latin_i s_i+\sum_{\alpha} \greek_\alpha \sigma_\alpha + \beta\sum_{i,\alpha}w_{i\alpha} C_{i\alpha}\right)}=\\
        &=\int_0^{+\infty}\prod_{i,\alpha} dw_{i\alpha}\,e^{- \sum_{i,\alpha}w_{i\alpha}\left(\beta C_{i\alpha}+\latin_i+\greek_\alpha\right)}=\\
        &=\prod_{i,\alpha}\int_0^{+\infty} dw_{i\alpha}\,e^{- w_{i\alpha}\left(\beta C_{i\alpha}+\latin_i+\greek_\alpha\right)}=\\
        &=\prod_{i,\alpha}\frac{1}{ \beta\, C_{i\alpha}+\latin_i+\greek_\alpha}
    \end{split}
\end{equation}
\noindent
This result lets us decompose $P_{\text{sub OT}}\,(G \!\mid\! \{t,\theta\})$ in edge by edge terms or as a product over the pairs $(i,\alpha)$:

\begin{equation}
\begin{split}
    P_{\text{sub OT}}\,(G \!\mid\! \beta, \{t,\theta\}) &= \prod_{i, \alpha} \left(\beta\, C_{i\alpha}+\latin_i+\greek_\alpha\right)\,\,e^{\,-w_{i\alpha}(G)\, \left(\beta\,C_{i\alpha} + \,\latin_i + \,\greek_\alpha\right)}\\
    &\equiv \prod_{i, \alpha} \underbrace{r_{i\alpha}\,e^{-r_{i\alpha}\,w_{i\alpha}(G)}}_{P\,\left(w_{i\alpha}\vert r_{i\alpha}\right)}    
\end{split}
\end{equation}

\noindent
where we define the rate parameter $r_{i\alpha} = \left(\beta\, C_{i\alpha}+\latin_i+\greek_\alpha\right)$. Thus, the weight of each edge $(i,\alpha)$ in the graph is an independent and exponentially distributed random variable $w_{i\alpha}$. Its probability density function is the conditional probability $P\left(w_{i\alpha}\vert r_{i\alpha}\right)$, with rate parameter $r_{i\alpha}$. When we compute the expected value $\langle w_{i\alpha}\rangle$ with respect to $P\left(w_{i\alpha}\vert r_{i\alpha}\right)$ we get $1/r_{i\alpha}$.

To determine the values of the Lagrange multipliers that enforce the strengths to match the empirical ones observed in a real graph \( G^* \), we solve the self-consistent equations obtained by maximizing the log-likelihood with respect to each parameter in the set \(\{t, \theta\}\):
\begin{eqnarray}
\nonumber
&&\mathcal{L}(G^* \!\mid\! \beta, \{t,\theta\}) = \log P(G^* \!\mid\! \beta, \{t,\theta\}) \\
&& = -\mathcal{H}(G^* \!\mid\! \{t,\theta\}) - \log Z\,(\beta, \{t,\theta\}) - \beta \,U(G^*)
\end{eqnarray}

\noindent
Since the energy term does not depend on any Lagrange multiplier, the ML condition states that:
\begin{eqnarray}
&&\frac{\partial \mathcal{L}(G^* \!\mid\! \beta, \{t,\theta\})}{\partial \theta_\alpha}\Bigg|_{\theta_\alpha = \theta_\alpha^*}\!\!\!\!\!\!\!\!\!
= \\
\nonumber
&&\left[ -\frac{\partial \mathcal H(G^* \!\mid\! \{t,\theta\})}{\partial \theta_\alpha} - \frac{1}{Z\,(\beta, \{t,\theta\})} \frac{\partial Z\,(\beta, \{t,\theta\})}{\partial \theta_\alpha} \right]_{\theta_\alpha = \theta_\alpha^*} \!\!\!\!\!\!\!\!\!=0
\end{eqnarray}

\begin{eqnarray}
&&\frac{\partial \mathcal{L}(G^* \!\mid\! \beta, \{t,\theta\})}{\partial t_i}\Bigg|_{t_i = t_i^*}\!\!\!\!\!\!\!\!\!
= \\
\nonumber
&&\left[ -\frac{\partial \mathcal H(G^* \!\mid\! \{t,\theta\})}{\partial t_i} - \frac{1}{Z\,(\beta, \{t,\theta\})} \frac{\partial Z\,(\beta, \{t,\theta\})}{\partial t_i} \right]_{t_i = t_i^*} \!\!\!\!\!\!\!\!\!=0
\end{eqnarray}

\noindent
As shown in Eq.(\ref{eq:dot_prod}), all terms coupled to the Lagrange multipliers that appear in $\mathcal H(G^* \!\mid\! \{t,\theta\})$ take the form of a dot product. So we obtain:

\begin{eqnarray}
\label{eq:sigma_<sigma>}
\frac{\partial \mathcal H(G^* \!\mid\! \{t,\theta\})}{\partial \theta_\alpha} \Bigg|_{\theta_\alpha = \theta_\alpha^*} \!\!\!\!\!\!\!\!\!\!\!\!\!\!\!\! &&= \sigma_\alpha(G^*) \\
\nonumber
&&= -\frac{1}{Z(\beta, \{t,\theta\})} \sum_{G} \sigma_\alpha(G) \,e^{-\mathcal H(G \mid \{t,\theta\})-\beta \,U(G)} = \langle \sigma_\alpha \rangle 
\end{eqnarray}

\begin{eqnarray}
\label{eq:s_<s>}
\frac{\partial \mathcal H(G^* \!\mid\! \{t,\theta\})}{\partial t_i} \Bigg|_{t_i = t_i^*}  \!\!\!\!\!\!\!\!\!\!\!\!\!\!\!\! &&= s_i(G^*) \\
\nonumber
&&= -\frac{1}{Z(\beta, \{t,\theta\})} \sum_{G} s_i(G) \,e^{-\mathcal H(G \mid \{t,\theta\})-\beta\,U(G)} = \langle s_i \rangle 
\end{eqnarray}

\noindent
Focusing on Eq.(\ref{eq:s_<s>}) solely, we replace the sum over $G$ by an integral over all $w_{j\gamma}$ and we get that:

\begin{equation}
\frac{1}{Z}\,
\int_{0}^{\infty}
\Bigl[\sum_{\alpha} w_{i\alpha}\Bigr]
\exp \Bigl[
-\!\sum_{j,\gamma}
w_{j\gamma}\,(\beta\,C_{j\gamma}+t_j+\theta_\gamma)
\Bigr]
\,\prod_{j,\gamma} dw_{j\gamma}
\end{equation}
\noindent
Because the exponential factorizes, 
\begin{equation}
\sum_{\alpha}\int_0^\infty \!\!\!\!
w_{i\alpha}
\prod_{j,\gamma}
\biggl[\,
e^{-\,(\beta\,C_{j\gamma}+t_j+\theta_{\gamma})\,w_{j\gamma}}
\,dw_{j\gamma}\biggr]
\end{equation}
\noindent  
we can rewrite the integral as:
\begin{equation}
\sum_\alpha \int_{0}^{\infty} \!\!\!\!
w_{i\alpha}\,
e^{-\,[\beta\,C_{i\alpha}+t_i+\theta_{\alpha}]\,w_{i\alpha}}
\,dw_{i\alpha} 
\prod_{\substack{(j,\gamma)\neq(i,\alpha)}}\! \int_{0}^{\infty} e^{-\,(\beta\,C_{j\gamma}+t_j+\theta_{\gamma})\,w_{j\gamma}} \,dw_{j\gamma}
\end{equation}
\noindent
Each $\alpha$ term in the integral over $i,\alpha$ gives as a result $1\,/\,(\beta C_{i\alpha} + t_i +\theta_\alpha)^2$ while the integral over each $j,\gamma$ term yields:
\begin{equation}
\prod_{\substack{(j,\gamma)\neq(i,\alpha)}}\!\frac{1}{\beta\,C_{j\gamma}+t_j+\theta_{\gamma}}\;\;=\;\frac{Z}{\frac{1}{\beta\,C_{i\alpha}+t_i+\theta_{\alpha}}}
\end{equation}
\noindent
Putting everything together and repeating such computations also for Eq.(\ref{eq:sigma_<sigma>}) we have that:
\begin{equation}
\label{eq:self_s}
s^*_i = \sum_{\alpha} \frac{1}{\beta\, C_{i\alpha} + \latin_i^* +  \greek_\alpha^*}  \qquad \forall i=1,\dots,N
\end{equation}

\begin{equation}
\label{eq:self_sigma}
\sigma^*_\alpha = \sum_{i} \frac{1}{\beta\, C_{i\alpha} +
\latin_i^* + \greek_\alpha^*} \qquad \forall \alpha=1,\dots,M
\end{equation}

\subsubsection*{Convexity of the optimization problem}
Solving Equations~(\ref{eq:self_s}) and (\ref{eq:self_sigma}) is equivalent to directly computing the gradient of the log-likelihood function while keeping the constraints fixed. This approach effectively transforms the problem into an optimization task. The convexity of the optimization problem described in Equation~(\ref{eq:free_en}) can be assessed by evaluating the Hessian matrix of $\mathcal{L}(G^* \!\mid\! \beta, \{t,\theta\})$. For the sake of simplicity, we will omit all dependencies of this function in the following discussion. The Hessian takes the form of a block matrix where each block contains negative second-order partial derivatives of $\mathcal{L}$ with respect to all possible combinations of the Lagrange multipliers:

\begin{eqnarray}
&&\frac{\partial^2 \mathcal L}{\partial t_{j}\,\partial t_{k}} = 
\\
&&
\nonumber
\frac{\partial}{\partial t_{j}}
\biggl(\sum_{\alpha}\,\frac{1}{\beta\,C_{k\alpha}+t_{k}+\theta_{\alpha}}
\biggr)
\;=\;
\begin{cases}
\displaystyle-\sum_{\alpha}\,\frac{1}{\bigl(\beta\,C_{k\alpha}+t_{k}+\theta_{\alpha}\bigr)^{2}}
&\text{if }j = k,
\\
0
&\text{if }j \neq k.
\end{cases}
\end{eqnarray}
Hence, in compact form,

\begin{equation}
\frac{\partial^2 \mathcal L}{\partial t_{j}\,\partial t_{k}}
\;=\;
-\,\delta_{jk}
\sum_{\alpha}\,\frac{1}{\bigl(\beta\,C_{k\alpha}+t_{k}+\theta_{\alpha}\bigr)^{2}}
\end{equation}

\noindent
In other words, the blocks of the Hessian matrix of \(\mathcal L\) corresponding to the second derivatives with respect to the Lagrange multipliers within the same layer are diagonal. Similarly, we have that:
\begin{align}
&\frac{\partial^2 \mathcal L}{\partial \theta_{\beta}\,\partial \theta_{\gamma}} =
-\,\delta_{\beta\gamma}\,\sum_{i}\;\frac{1}{\bigl(\beta\,C_{i\beta}+t_{i}+\theta_{\beta}\bigr)^{2}}\\
&\frac{\partial^2 \mathcal L}{\partial t_{j}\,\partial \theta_{\gamma}} = 
-\,\frac{1}{\bigl(\beta\,C_{j\gamma}+t_{j}+\theta_{\gamma}\bigr)^{2}}
\end{align}

\subsection*{Implementation of the numerical solution}
Here we show how to implements an optimization procedure to determine the Lagrange multipliers \( t \) and \( \theta \) by directly optimizing the log-likelihood function rather than explicitly solving the self-consistent equations. The key idea is to leverage gradient-based optimization to find the optimal values of \( t \) and \( \theta \) that enforce the expected strength constraints. The metrics \texttt{perc\_error} is defined to track how well the row and column constraints are satisfied: it computes the percentage of "mass" of the graph that misses the strength constraint. Instead of explicitly solving the self-consistent Equations~(\ref{eq:self_sigma}) and (\ref{eq:self_s}) for \( t \) and \( \theta \), the approach minimizes the negative of the log-likelihood 

\begin{equation}
\nonumber
\mathcal{L}(G^* \!\mid\! \beta, \{t,\theta\}) = 
\sum_{i,\alpha} w_{i\alpha}(G^*)\, \left(\beta \,C_{i\alpha} + \latin_i + \greek_\alpha\right) - \sum_{i,\alpha}\log\,\left(\beta\, C_{i\alpha}+\latin_i+\greek_\alpha\right)
\end{equation}

\noindent
via Stochastic Gradient Descent (SGD). The log-likelihood function serves as the loss function, whose gradients are computed using PyTorch \cite{NEURIPS2019_bdbca288} automatic differentiation, and the parameters \( t \) and \( \theta \) are updated iteratively. In this procedure, each \( t_i \) and \( \theta_\alpha \) are initialized as \( 2/s^*_i \) and \( 2/\sigma^*_\alpha \), respectively, and the learning rate is set high at the outset to allow for faster convergence, then it gradually decreases when the Loss function increases instead. The optimization continues up to a maximum number of iterations, unless the error falls below \( 10^{-3} \), in which case it terminates early. If the solution has not converged after the specified number of steps, the process halts and saves the current values of $t$ and $\theta$ to start a new optimization. This gradient-based method avoids explicitly inverting equations and provides a straightforward means to enforce the required constraints on the row and column sums by allowing the optimizer to drive the system toward the correct parameters \( t \) and \( \theta \).

\newpage
\begin{algorithm}[ht]
\caption{Gradient-Based Optimization of $t$ and $\theta$}
\label{alg:opt_grad_L}
\begin{algorithmic}[1]

\Require 
  A set of matrix dimensions \texttt{dims}; 
  coupling costant range \texttt{betas}; 
  maximum number of steps \texttt{num\_steps}; 
  initial learning rate \texttt{lr0}; 
  momentum for SGD \texttt{m0};
  maximum weight \texttt{maxweigth}.

\Ensure 
  Optimized parameters $t$, $\theta$ stored for each \texttt{dim} and each $\beta$.

\Statex
\For{\texttt{dim} \textbf{in} \texttt{dims}}
    \State \textbf{generate} \texttt{betas}
    \State \textbf{generate} cost matrix \texttt{cost} as random in $[0,1)$
    \State \textbf{transform} \texttt{cost} \Comment{when required}
    \State \textbf{generate} a random matrix \texttt{M} in $[0,\texttt{maxweigth}]$ of shape (\texttt{dim},\texttt{dim}) \Comment{to compute the strengths}
    \State $r \gets$ row-sums of \texttt{M}
    \State $c \gets$ column-sums of \texttt{M}
    \State $t \gets 2/r$ \quad \Comment{Element-wise division}
    \State $\theta \gets 2/c$ \quad \Comment{Element-wise division}

    \For{\texttt{beta} \textbf{in} \texttt{betas}}
        \State \textbf{set up optimizer} \texttt{SGD} with \texttt{lr} $\leftarrow \texttt{lr0}$, momentum $\leftarrow \texttt{m0}$
        \State \textbf{initialize} \texttt{increase}$\leftarrow 0$, \texttt{patience}$\leftarrow 10$, \texttt{last\_loss}$\leftarrow \infty$
        \For{\texttt{step} $\gets 0$ \textbf{to} \texttt{num\_steps}}
            \State \textbf{optimizer.zero\_grad()}
            \State $\texttt{loss} \gets \mathcal L (t, \theta,\, r,\, c,\, \texttt{cost},\, \texttt{beta})$
            \State \textbf{loss.backward()}
            \State \textbf{optimizer.step()}
            \State \texttt{perc\_err} $\gets \texttt{perc\_error}(\texttt{beta},\, \texttt{cost},\, t,\, \theta,\, d,\, u)$
            
            \If{\texttt{loss} $>$ \texttt{last\_loss}}
                \State \texttt{increase} $\gets \texttt{increase} + 1$
            \EndIf
            \State \texttt{last\_loss} $\gets \texttt{loss}$
            
            \If{\texttt{increase} $\geq$ \texttt{patience}}
                \State \textbf{reduce learning rate} by factor $0.9$
                \State \texttt{increase} $\gets 0$
            \EndIf

            \If{\texttt{perc\_err} $< 10^{-3}$}
                \State \textbf{save} $(t,\theta)$ 
                \State \textbf{break}
            \EndIf

            \If{\texttt{step} $=$ \texttt{num\_steps}}
                \State \textbf{save} $(t,\theta)$ \Comment{to start a new opt from the current $t$ and $\theta$}
                \State \textbf{break}
            \EndIf
        \EndFor
    \EndFor
\EndFor

\end{algorithmic}
\end{algorithm}

\newpage
\section*{Data availability}
The datasets generated and analyzed during the current study are available from the corresponding author upon reasonable request. 

\section*{Code availability}
The codes developed and used for the simulations and analyses presented in this study are available from the corresponding author upon reasonable request. 

\section*{Acknowledgements}
D.M., A.P. and R.P. acknowledge the financial support under the National Recovery and Resilience Plan (NRRP), Mission 4, Component 2, Investment 1.1, Call for tender No. 104 published on 2.2.2022 by the Italian Ministry of University and Research (MUR), funded by the European Union – NextGenerationEU– Project Title  "WECARE – WEaving Complexity And the gReen Economy" – CUP 20223W2JKJ by the Italian Ministry of Ministry of University and Research (MUR).

\section*{Author contributions}
L.B., D.M., and R.P. carried out the numerical simulations. L.B., F.S., and A.P. developed the analytical framework. The analysis was designed by L.B., D.M., A.P., G.C., and F.S. All authors contributed equally to the writing and revision of the manuscript.

\section*{Competing interests}
The authors declare no competing interests.

\end{document}